# Machine learning via artificial neural networks coupled with density functional theory and experiments for thermodynamic optimization of high-entropy alloys for hydrogen storage at room temperature

Shivam Dangwal[1,2], Pranav Kumar[3], Yuji Ikeda[3], Blazej Grabowski[3] and Kaveh Edalati[1,2,*]

[1] WPI, International Institute for Carbon-Neutral Energy Research (WPI-I2CNER), Kyushu University, Fukuoka, Japan
[2] Department of Automotive Science, Graduate School of Integrated Frontier Sciences, Kyushu University, Fukuoka, Japan
[3] Institute for Materials Science, University of Stuttgart, Stuttgart, Germany

*Corresponding author (E-mail: kaveh.edalati@kyudai.jp; Tel/Fax: +81 92 802 6744)

High-entropy alloys (HEAs) have received considerable attention for hydrogen storage because of their compositional flexibility; however, designing HEAs with optimal thermodynamics is critical. This study employs machine learning via artificial neural networks (ANN) and density functional theory (DFT) to design a novel AB-type $Ti_xNb_{2-x}VCrMnFe$ ($x$ = 0.5−2.0) high-entropy system for hydrogen storage at ambient temperature (A: Ti, V and Nb, and B: Cr, Mn and Fe). Both ANN and DFT predict that the hydride formation enthalpy decreases to negative values with increasing the titanium content. Two alloys with $x > 1.5$ are predicted to achieve enthalpies within the −25 to −39 kJ/mol range, making them appropriate for room-temperature hydrogen storage. Experiments demonstrate good agreement with the enthalpy predictions, with the Ti-rich alloys showing reversible hydrogen storage with fast kinetics at room temperature. These results provide a framework for reliable use of data analysis and *ab initio* calculations to explore high-entropy hydrides as hydrogen storage materials.

## 1. Introduction

A global shift towards renewable energy, particularly for electrical energy, has been observed in this decade to reduce carbon emissions [1]. In industries where complete electrification is not feasible, hydrogen can act as a green energy carrier [1,2]. One kilogram of hydrogen carries around 120 MJ of energy (excluding heat from produced water) to 142 MJ (including heat from produced water), and this exceeds the energy carried by one kilogram of any other chemical fuel [1,3]. However, the energy carried by one cubic meter of hydrogen at atmospheric pressure is around 9.8 MJ, which is much less than the levels set by different organizations, including the United States Department of Energy [4]. The volumetric energy density of hydrogen can be increased by solid-state storage, e.g., in the metal hydride form [5,6].

Storing hydrogen in solid form offers a compact and safe storage strategy; however, most metal hydrides face kinetic (difficult activation) and thermodynamic (high-temperature desorption) issues [5,6]. The kinetics of hydrogen storage can be boosted through several methods, such as the addition of catalysts [6,7], applying ball milling [8,9], inducing severe plastic deformation [10-12], introducing interphase boundaries [13], etc. However, despite significant efforts, the optimization of the kinetic and thermodynamic properties remains significantly challenging. The hydride formation enthalpy for storing hydrogen at room temperature (298−303 K) under plateau pressures of 0.1-3 MPa should be within −25 to −39 kJ/mol, by assuming a hydride formation entropy of −130 to −110 J/mol/K [14,15]. An integration of dedicated A- and B-type elements (A: strong hydrogen affinity, B: weak affinity for hydrogen) [16,17] within the crystal lattice can, in principle, adjust the hydride formation enthalpy to lie within the required range for room-temperature storage. However, typically, phase competitions do not allow for easy compositional changes because the added elements may form undesirable secondary or ternary phases instead of being incorporated into the crystal structure of the storing media. Regarding this, the concept of high-entropy alloys (HEAs) has provided new opportunities to develop materials with high compositional flexibility within a single phase [18].

Alloys with five or more principal elements and an entropy over 1.5$R$ ($R$ = 8.314 J/mol/K) are termed HEAs [18,19]. Recent studies demonstrate the potential application of HEAs in diverse fields, such as in Ni-metal hydride batteries [20,21], refractory materials [22], biomaterials [23], and hydrogen storage materials [14]. High configurational entropy and consequently low Gibbs energy in these alloys are the main factors that enable HEAs to exist with a smaller number of phases and excellent compositional tunability [24]. As a result, HEAs offer the potential to achieve desirable phases with optimized elemental combinations for effective solid-state hydrogen storage. Recent studies showed that a few HEAs can exhibit hydrogen absorbing/desorbing behavior–at ambient temperature without any activation, including Ti-Zr-Cr-Mn-Fe-Ni [15,16,25], Ti-V-Zr-Cr-Mn-Fe-Ni [13,19], Ti-Zr-Fe-Ni [26] and Zr-Nb-Fe-Co [27]. However, some HEAs exhibit high storage capacities up to 3.8 wt%, but they still need elevated temperatures for desorbing hydrogen, such as V-Ti-Cr-Fe-Mn [28], Ti-V-Zr-Hf-Nb [29], Ti-V-Zr-Nb-Ta [30] and Ti-Zr-Nb-Hf-Ta [31]. Therefore, designing HEAs with proper thermodynamics (enthalpy and entropy) for room-temperature hydrogen storage is still a critical issue. The feasibility of a material to store hydrogen at ambient temperature is primarily determined by its enthalpy. In contrast, the entropy variation in hydrogenation mainly arises from the transformation of gaseous hydrogen to atomic hydrogen within the alloy lattice and therefore remains reasonably similar across different materials (−130 to −110 J/mol/K) [32-34].

The formation enthalpy of hydrides can be determined either by experiments or theoretical calculations. One way to experimentally determine the formation enthalpy is by measuring the equilibrium plateau pressures from pressure-composition-temperature (PCT) isotherms at various temperature levels and constructing the van't Hoff plot [15,35]. Another way to determine the formation enthalpy is by using differential scanning calorimetry (DSC). Theoretically, the formation enthalpy can also be calculated using first-principles calculations such as density functional theory (DFT) [36], as demonstrated in several previous studies to evaluate various hydrides for hydrogen storage [25,37,38]. Although both experiments and theoretical methods can provide the formation enthalpy of hydrides, experiments require a lot of time and financial resources, whereas DFT requires vast computational resources. Machine learning (ML) models, which are trained on available experimental or theoretical data, provide a complementary, rapid, and cost-effective approach to predicting thermodynamic properties like enthalpy [15,32].

The potential of ML is well-established in various fields such as medical imaging [38], agriculture [40], finance [41], computational chemistry [42], etc. In materials science, ML can be employed to design and analyze materials and for inverse design [43]. Recent progress in integrating large language models (LLMs) has eased the transformation of ideas into practical tools and simplified the interpretation of high-dimensional data [44]. In the hydrogen storage research, different ML techniques have been used for material discovery (Bayesian optimization [45], and support vector machine and gradient boosted decision tree [46]), predicting thermodynamic properties (Gaussian process regression (GPR) [15,47], and random forest [32]), storage capacity estimation (multi-layer perceptron, random forest, support vector machine, and gradient boost [48]) and hydride classification (neural networks [49]). Particularly, ML through artificial neural networks (ANN) has exhibited high potential in analyzing complex data, which are difficult for the human brain to comprehend [40], and thus, such ML-ANN analyses have high potential for the hydrogen storage field.

The objectives of the current investigation are to employ ML and discover new HEAs for hydrogen storage at ambient temperature. Five novel AB-type HEAs with composition $Ti_xNb_{2-x}VCrMnFe$ ($x$ = 0.5, 1.0, 1.5, 1.8, 2.0) are considered, and the impact of changing the ratio of titanium (with small atomic size) to niobium (with large atomic size) on hydrogen storage properties is investigated. It is noted that here AB-type HEA refers to an equal fraction of hydride-forming and non-hydrogen-forming elements in the HEAs. Although these materials have an ordered C14 structure, they are referred to as HEAs, as this terminology is widely used for intermetallics and even ceramics in recent years [18]. In these HEAs, titanium, vanadium and niobium contribute to the formation of hydride, and chromium, manganese and iron lower the hydrogen desorption temperature. In addition to the introduction of the senary Ti-Nb-V-Cr-Mn-Fe system, a main novelty of this study lies in the systematic study of Ti/Nb ratio on hydrogen storage properties of the senary HEAs using experiments, ML and DFT. The formation enthalpy is calculated using ML through ANN, while the calculations are verified with DFT and experimental results. Both DFT and experimental results verify the precision of the ML method to predict the thermodynamic behavior of high-entropy hydrogen storage materials.

## 2. Methodology

### 2.1 Experiment

Vacuum arc-melting was used to produce the ingots of $Ti_xNb_{2-x}VCrMnFe$ ($x$ = 0.5, 1.0, 1.5, 1.8, 2.0) by melting high-purity metallic chunks of chromium (99.99 %), iron (99.9 %), manganese (99.9 %), niobium (99.9 %), titanium (99.9 %) and vanadium (99.7 %). To prevent oxidation during synthesis, an inert atmosphere of argon was used. The chemical homogeneity of the HEAs was ensured by remelting and rotating the ingots ten times during the synthesis process, as reported in previous studies [13,19]. The ingots after arc melting were cut using electric discharge machining and examined by different methods.

The crystal structure of the HEAs prior and after hydrogen absorption was identified using X-ray diffraction (XRD) with Cu Kα. Lattice parameters and peak positions were identified using Rietveld refinement analysis, with the help of the PDXL software.

The microstructural analysis of the HEAs was performed employing scanning electron microscopy (SEM) operated at 15 kV. Specimens for SEM examination were prepared by sectioning the ingots into cylindrical shapes (0.8 mm height and 5 mm radius). Surface preparation involved sequential grinding with emery sandpapers (grits: 400, 800, 1200 and 2000), fine polishing with diamond suspensions (particle sizes: 9 and 3 μm), and final surface finishing with a buff and colloidal silica (particle size: 60 nm). Compositional and phase analyses were performed using energy-dispersive X-ray spectroscopy (EDS) and electron backscatter diffraction (EBSD), respectively.

The nanostructure of the HEAs was analyzed using a transmission electron microscope (TEM). The TEM voltage was set to 200 kV. TEM samples were prepared by dispersing the ingot pieces in ethyl alcohol and pouring them onto carbon grids. High-resolution TEM θimages and fast Fourier transform (FFT) were carried out to analyze the local crystal structure.

Hydrogen storage performance was examined by a Sievert-type equipment. Pieces of ingot with 0.5 g mass were crushed and sieved with a sieve of 75-micrometer mesh size in an air atmosphere. Hydrogenation of the HEAs was initially done at 303 K without any thermal activation, but no hydrogenation was detected. The samples were then activated in vacuum at 673 K for 60 minutes, followed by three hydriding and dehydriding cycles at 303 K to obtain the PCT isotherms. Due to the high plateau pressures exhibited by the alloys, PCT isotherms could not be obtained at elevated temperatures, thereby limiting the feasibility of performing van't Hoff analysis. However, the enthalpy of hydride formation was estimated using the following thermodynamic relation [3,15,35]:

$$\ln \frac{P_{\mathrm{eq}}}{P_0} = \frac{\Delta H}{RT} - \frac{\Delta S}{R} \tag{1}$$

where $P_{\mathrm{eq}}$ stands for equilibrium pressure; $P_0$ stands for atmospheric pressure; $\Delta H$ stands for formation enthalpy; $\Delta S$ stands for the change in entropy, which was assumed to be −110 J/mol/K [32,34]; $T$ is the temperature of the hydrogenation/dehydrogenation reaction (303 K in this study), and $R$ stands for the gas constant. $P_{\mathrm{eq}}$ was obtained by numerical differentiation of the PCT isotherms, and the hydrogen pressure at the mid-point of the region with a slope close to zero in the derivative plot was taken as $P_{\mathrm{eq}}$. It should be noted that the entropy assumption of −110 J/mol/K which is based on reported experimental data [32,34], compared to the entropy of -130 J/mol/K which is the theoretical value for the ideal hydrogen to hydride transformation, generates a free energy difference ($-T\Delta S$) of only 6 kJ/mol at ambient temperature (298−303 K), which is not a significant value compared to the contribution of enthalpy to the free energy. Kinetic tests were also performed at 303 K for 10 minutes after the third cycle of the PCT analysis, followed by XRD analysis of the hydrogenated samples.

### 2.2 Machine Learning

A set of data containing alloy compositions and hydride formation enthalpies for 511 data points, as shown in the Appendix, was gathered for ML analysis. The input parameters for the neural networks were the atomic fractions of 26 elements, aluminum (Al), boron (B), carbon (C), cerium (Ce), chromium (Cr), cobalt (Co), copper (Cu), gadolinium (Gd), hafnium (Hf), holmium (Ho), iron (Fe), lanthanum (La), lithium (Li), magnesium (Mg), manganese (Mn), molybdenum (Mo), nickel (Ni), niobium (Nb), silicon (Si), tantalum (Ta), tin (Sn), titanium (Ti), tungsten (W), vanadium (V), yttrium (Y), and zirconium (Zr). All compositions were normalized to ensure that the sum of the atomic fractions of all constituent elements in each alloy equals 1. The target variable to be predicted was the formation enthalpy of the hydride in kJ/mol. When the formation enthalpy was not reported in the literature, it was calculated through the provided desorption PCT isotherms using Eq. 1, assuming the change in entropy is −110 J/mol/K [32-34]. Moreover, an additional model was trained using a refined dataset, which did not include the data points where −110 J/mol/K was assumed as a change in entropy value. Training of the dataset was done using ANN, following a procedure schematically shown in Fig. 1(a).

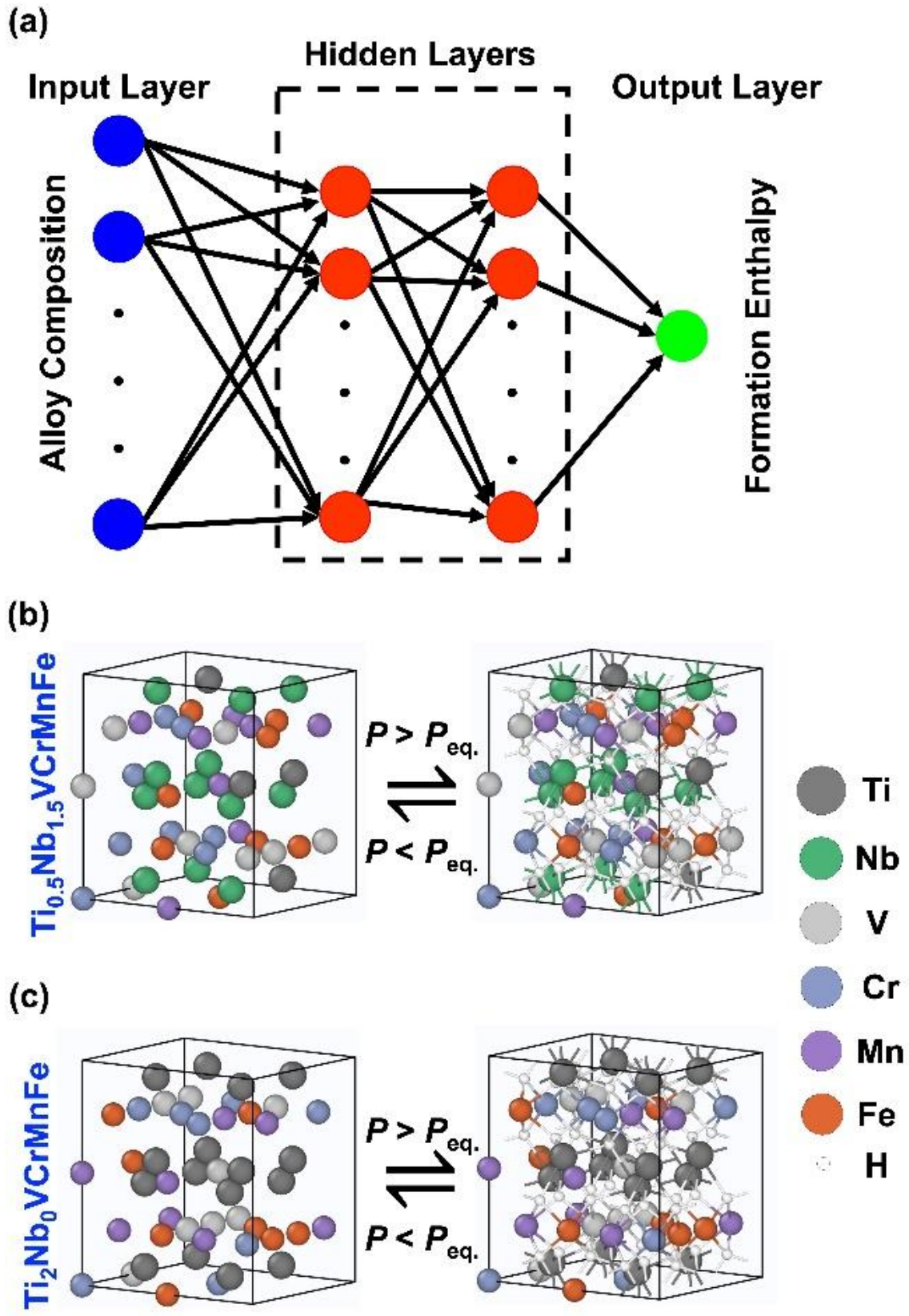

Fig. 1. (a) Schematic of the utilized artificial neural networks (ANN), in which the alloy composition is the input layer, the formation enthalpy is the target variable, and hidden layers consist of neurons, which enable the ANN to learn complex patterns in the dataset. (b, c) Optimized supercells for density functional theory (DFT) calculations of (b) $Ti_{0.5}Nb_{1.5}VCrMnFe$ and (c) $Ti_2Nb_0VCrMnFe$ high-entropy alloys before and after hydrogenation ($P$: hydrogen pressure, $P_{eq}$: equilibrium plateau pressure).

For developing the ANN model, MATLAB was used for training and testing the composition and enthalpy of HEAs. The reproducibility of the results was ensured by setting a random seed for a pseudo-random-number generator before all processes of training and validation. Before training, the outliers of the data were eliminated. To find the most appropriate way to eliminate outliers for ANN modeling, two different ways were examined. In the first method, median absolute deviation (MAD) residual analysis was used to statistically eliminate outliers. MAD was chosen to eliminate outliers instead of the Z-score and the standard deviation method, because the latter methods are very sensitive to extreme data points. MAD relies on the median, hence allowing it to effectively identify genuine outliers [50,51]. The residual ($r_i$) was calculated using a preliminary neural network,

$$r_i = y_i - \hat{y}_i \tag{2}$$

where $y_i$ stands for true enthalpy and $\hat{y}_i$ stands for the predicted enthalpy of the $i$-th alloy from the preliminary ANN model. The MAD was calculated using [51,52].

$$MAD = 1.4826 \times \text{median}(\lceil r_i - \text{median}(r_i) \rceil). \tag{3}$$

Assuming the residuals approximately follow a normal distribution, the MAD was scaled by a factor of 1.4826 to provide a consistent estimate of the standard deviation under normality [53]. Data points that did not satisfy Eq. 4 were considered as outliers.

$$|r_i - \text{median}(r_i)| \leq 3 \times \text{MAD}. \tag{4}$$

Fig. 2 shows a visual representation of the data points. A scatter plot of the residuals with outliers and inliers, detected from the MAD method, is shown in Fig. 2(a). The outlier threshold (at ±3 × MAD) is shown by a dashed line separating inliers and outliers. Fig. 2(b) shows a boxplot of the residuals. The symmetric position of the median in the boxplot indicates negligible skewness [54]. The outliers indicated by the MAD method lie beyond the 1.5 × interquartile range of the residuals. A hybrid method was also used to remove outliers, in which data points were first filtered using domain knowledge and then further refined through statistical outlier detection. The alloys in the dataset having a major C14 Laves phase were taken, and then outliers from those were statistically removed, similar to the method described above. After this process, two sets of data were available for training: (i) data after statistical outlier removal (476 data points), and (ii) data after hybrid outlier removal (234 data points). It should be noted that hybrid outlier removal leads to a substantial reduction in data points, which may affect the statistical power of the model. To ensure robust uncertainty quantification, a twenty-member bootstrap ensemble of ANN was employed.

Statistical metrics, coefficient of determination ($R^2$) and root mean squared error (RMSE) were employed as tools for evaluating the performance of ANN [15].

$$\text{Coefficient of Determination: } R^2 = 1 - \frac{\sum_{i=1}^{n}(\hat{y}_i - y_i)^2}{\sum_{i=1}^{n}(y_i - \overline{y})^2} \tag{5}$$

$$\text{Root Mean Squared Error: (RMSE)} = \sqrt{\frac{\sum_{i=1}^{n}(\hat{y}_i - y_i)^2}{n}} \tag{6}$$

where $y_i$ stands for the true enthalpy, $\hat{y}_i$ stands for the predicted enthalpy of the *i*-th alloy, $\bar{y}$ stands for the mean enthalpy, and *n* stands for the number of data points. Hyperparameters of the model were optimized with RMSE as the key metric. A detailed description of the ANN architecture is provided in the supplementary materials, along with a flowchart in Fig. S2. The validation of data was done by two methods: *K*-fold cross-validation with *K* equal to 5 and resubstitution validation, as attempted earlier in [15]. The *K*-fold cross-validation was chosen because it shows higher accuracy and a more reliable estimate of the performance of the model than holdout validation [55].

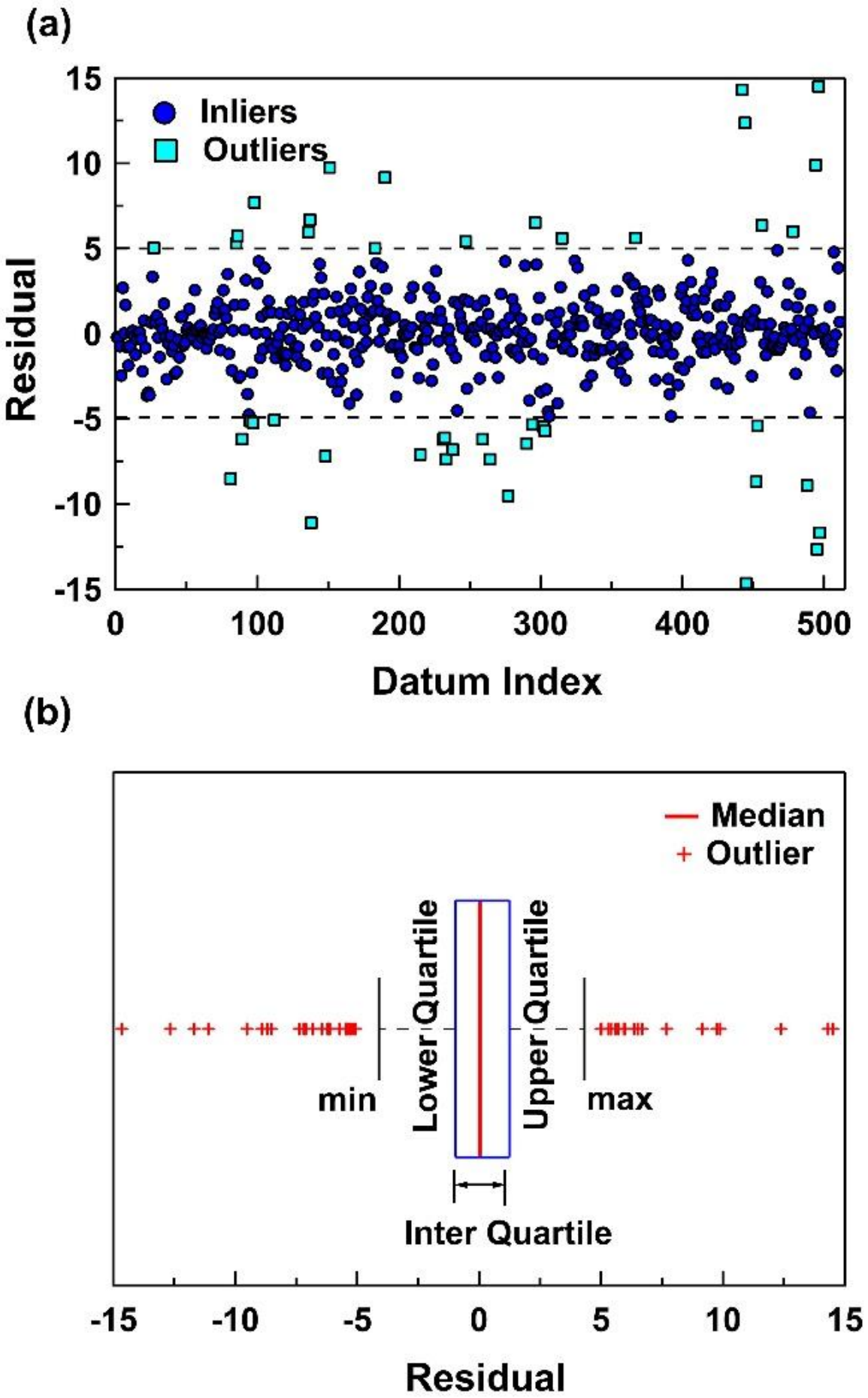

Fig. 2. Visualization of inlier and outlier data for machine learning. (a) Scatter plot of residuals indicating outliers and inliers. The dashed lines represent the outlier threshold, indicating that 7 % of the data are outliers. (b) Boxplot of the residuals. The median is shown as the red line. The outliers indicated by the median absolute deviation (MAD) method are shown by red plus symbols.

To prevent data leakage, identical compositions were grouped together and assigned exclusively to either the training or the testing data set. To understand and visualize the effect of elements on the formation enthalpy, partial dependence plots (PDP) and elemental sensitivity analysis were conducted for the target elements Ti, V, Nb, Cr, Mn and Fe, as these elements are present in the target HEA system $Ti_xNb_{2-x}VCrMnFe$ ($x$ = 0.5, 1.0, 1.5, 1.8, 2.0). For the PDPs, the concentration of alloys was varied between minimum and maximum values, and then the predicted enthalpy was calculated using the developed ANN. This variation in composition was plotted against the predicted enthalpy as PDP [56]. For element sensitivity analysis, the target elements (Ti, V, Nb, Cr, Mn and Fe) in composition were set to zero, and then the new formation enthalpy was predicted. Additionally, the change between the original prediction and the new prediction was calculated, which is described as the impact score. To improve the composition-formation enthalpy predictions, another model was also developed using 13 possibly important descriptors in hydrogen storage materials, including atomic weight, maximum difference in atomic weights, mean absolute deviation of atomic weight, mean covalent radius, mean absolute deviation of covalent radius, mean electronegativity, maximum difference in electronegativity, valence electron concentration, fraction of valence electron in *d*-shell, mean ionic character, number of components, melting point and mean absolute deviation of melting temperature.

### 2.3 *Ab initio* Calculations

$Ti_xNb_{2-x}VCrMnFe$ HEAs ($x$ = 0.0, 0.5, 1.0, 1.5, 2.0) and corresponding high-entropy hydrides were modeled using special quasirandom structures (SQSs) [57], generated using the icet package [58]. The structural models were based on a 2 × 2 × 1 supercell of the $AB_2$-type C14 Laves phase unit cell, resulting in configurations with 48 atoms. It should be noted that the presence of minor amounts of body-centered cubic (BCC) phase or short-range ordering was not considered in generating the 2 × 2 × 1 supercells using SQS. Atomic species were assigned such that Ti and Nb, with larger atomic sizes, occupied the A-type sublattice, while V, Cr, Mn, and Fe, with smaller atomic sizes, occupied the B-type sublattice. Although V works chemically as a hydride-forming element, structurally, it occupies B-sites [59]. This atomic arrangement, which also led to the best consistency with experimental diffraction data, was used for the calculations. The correlation functions of the nearest-neighbor doublet, triplet, and quartet clusters within cutoff distances of 6.0, 3.0, and 3.0 Å, respectively, were optimized using a simulated annealing algorithm to approximate fully random alloys. As an example, the SQS models of $Ti_{0.5}Nb_{1.5}VCrMnFe$ and-$Ti_2Nb_0VCrMnFe$ HEAs and their corresponding hydrides are displayed in Fig. 1(b) and Fig. 1(c), respectively.

An $AB_2$-type Laves phase alloy has 17 tetrahedral interstitial sites in its formula unit, which may incorporate hydrogen [60]. These sites are categorized according to their local atomic: (i) sites organized with four B atoms ($B_4$), sites organized with one A and three B atoms ($AB_3$), and (ii) sites organized with two A and two B atoms ($A_2B_2$). Energetically preferred sites for hydrogen incorporation are the $A_2B_2$ sites, which have been shown by various *ab initio* studies on conventional Laves phases [59,61–63]. Further, in C14 hexagonal Laves phases, four symmetrically distinct types of tetrahedral $A_2B_2$ interstitial sites are present [64]. Among these, the $12k_2$ interstitial tetrahedra are unique because they do not share a common face up to full occupancy to form $AB_2H_3$ [60]. Previous DFT studies on the $TiCr_2$ Laves phase [59] have demonstrated a linear correlation between cumulative hydrogen binding energy and hydrogen concentration at the $12k_2$ sites at least up to $TiCr_2H$ composition, indicating minimal H–H interactions among these sites. Therefore, restricting hydrogen occupancy to the $12k_2$ sites serves as an effective strategy to minimize computational cost, a method also adopted in prior work [25], and applied in the present study.

*Ab initio* calculations were conducted with plane-wave basis sets implemented in the VASP software [65,66]. The core-valence electron interactions were modeled using the projector augmented-wave (PAW) method [67], and the generalized gradient approximation (GGA) in the Perdew–Burke–Ernzerhof (PBE) form [68] was used to treat exchange–correlation effects. Spin-polarization was considered for all the calculations. The Brillouin zone was sampled on a Γ-centered $4 \times 4 \times 4$ *k*-point grid in conjunction with the Methfessel–Paxton smearing with 0.1 eV width. A plane-wave cutoff energy of 400 eV was used. Electronic convergence was achieved with $1 \times 10^{-5}$ eV energy tolerance. The structures were optimized using the conjugate-gradient method [69] with relaxing cell parameters and atomic coordinates, achieving residual atomic forces below $1 \times 10^{-3}$ eV/Å. For each alloy, 100 SQS configurations were screened using Γ-point sampling. From these 100 configurations, three representatives corresponding to maximum, minimum and average formation enthalpy were selected, which were further subjected to structural relaxation using a refined *k*-point grid to establish the formation enthalpies and corresponding error bars.

## 3. Results

### 3.1 Experiments

Fig. 3 shows PCT isotherms for three cycles after thermal activation, and Table 1 gives the average hydrogen storage capacity and hydrogen-to-metal ratio (H/M) for the $Ti_xNb_{2-x}VCrMnFe$ ($x$ = 0.5, 1.0, 1.5, 1.8, 2.0) alloys. The HEAs with $x$ = 0.5, 1.0, and 1.5 did not achieve complete hydrogenation because the required hydrogen pressures exceeded the operational limits of the experimental apparatus. On the other hand, HEAs with $x$ = 1.8 and 2.0 approached an H/M ratio of 1, indicating that they nearly reached their maximum capacity. The plateau pressure for the HEAs with $x$ = 0.5, 1.0, and 1.5 could not be measured, and thus, their hydride formation enthalpy remains unclear. However, formation enthalpies calculated using Eq. 1 are −24.9 kJ/mol for $Ti_{1.8}Nb_{0.2}VCrMnFe$ and −25.1 kJ/mol for $Ti_2Nb_0VCrMnFe$ by assuming an entropy of −110 J/mol/K. It should be noted that PCT measurements at two different temperatures and the corresponding van`t Hoff analyses for $Ti_{1.8}Nb_{0.2}VCrMnFe$ and $Ti_2Nb_0VCrMnFe$ give enthalpy values close to these two values, as shown in Fig. S1. However, since only two close temperatures of 303 and 313 K could be used (due to the pressure limits in the gas absorption machine), these van`t Hoff analyses are not experimentally considered so reliable. The calculated formation enthalpies −24.9 and −25.1 kJ/mol for these two alloys lie within the desirable limits for room-temperature hydrogen storage (−25 to −39 kJ/mol) [14,15]. Fig. 3(f) shows the third PCT cycles for all HEAs. With rising Ti content, the plateau pressure decreases, and hence, the formation enthalpy becomes more negative, consistent with trends reported in Ti-Nb-based systems [70]. Hydrogenation kinetic curves for the HEAs $Ti_{1.8}Nb_{0.2}VCrMnFe$ and $Ti_2Nb_0VCrMnFe$ are shown in Fig. 4(a,b). Both HEAs show fast absorption kinetics, achieving 90 % of maximum capacity in less than 2 minutes, a behavior consistent with that reported for other HEAs [71-73]. Hydrogenation cyclic test was conducted on HEA $Ti_2Nb_0VCrMnFe$ for 20 cycles as shown in Fig. 4(c). A similar hydrogen storage capacity is retained after 20 cycles, indicating good cyclability of HEA $Ti_2Nb_0VCrMnFe$. PCT isotherm for the 21st cycle, shown in Fig. 3(e) for HEA $Ti_2Nb_0VCrMnFe$, also exhibits a similar curves as initial cycles. These results confirm that, in addition to good thermodynamics and kinetics performance, the material has good cycling performance, although its weakness is the need for an initial thermal activation process.

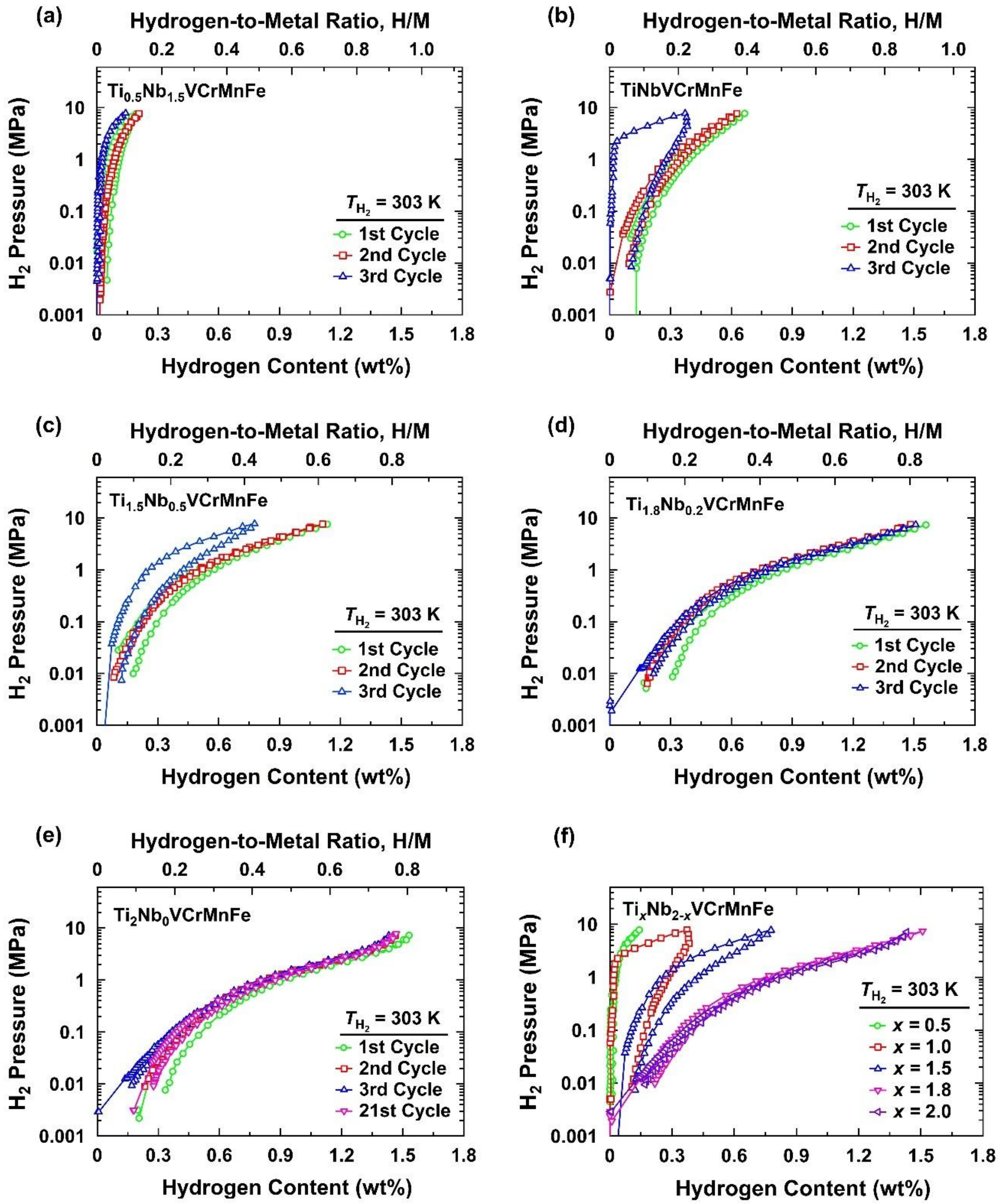

Fig. 3. Room-temperature hydrogen storage performance of $Ti_xNb_{2-x}VCrMnFe$ with low niobium content. Pressure-composition-temperature (PCT) isotherms at room temperature for (a) $Ti_{0.5}Nb_{1.5}VCrMnFe$, (b) TiNbVCrMnFe, (c) $Ti_{1.5}Nb_{0.5}VCrMnFe$, (d) $Ti_{1.8}Nb_{0.2}VCrMnFe$ and (e) $Ti_2Nb_0VCrMnFe$. (f) Third-cycle PCT isotherms of all high-entropy alloys, showing that with increasing titanium content, plateau pressure decreases.

Table 1. Hydrogen storage capacity and H/M ratio for the high-entropy alloy system $Ti_xNb_{2-x}VCrMnFe$ ($x$ = 0.5, 1.0, 1.5, 1.8, 2.0).

| High-entropy alloy | H/M ratio | Storage capacity (wt%) |
|---|---|---|
| $Ti_{0.5}Nb_{1.5}VCrMnFe$ | 0.1 | 0.2 |
| TiNbVCrMnFe | 0.3 | 0.6 |
| $Ti_{1.5}Nb_{0.5}VCrMnFe$ | 0.5 | 1.0 |
| $Ti_{1.8}Nb_{0.2}VCrMnFe$ | 0.8 | 1.5 |
| $Ti_2Nb_0VCrMnFe$ | 0.8 | 1.5 |

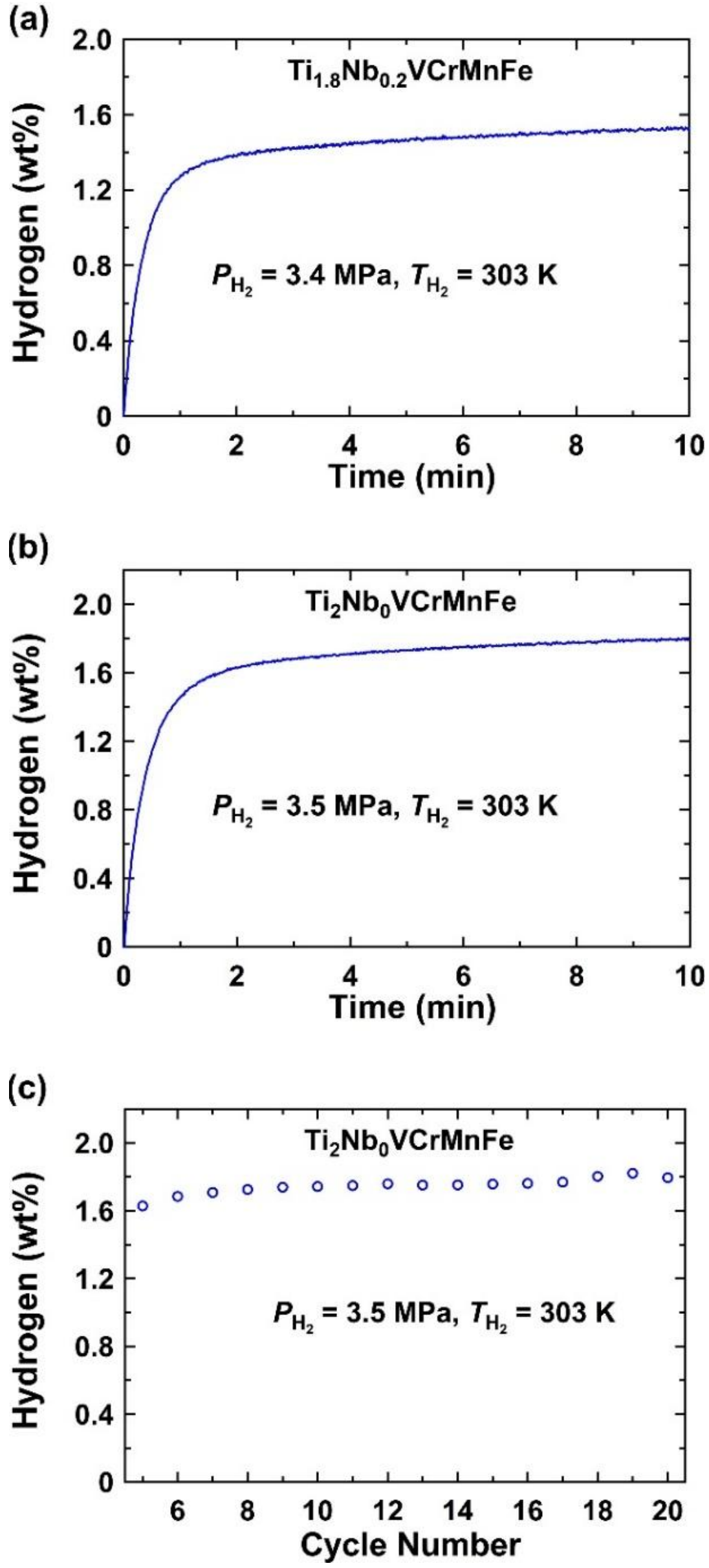

Fig. 4. Fast hydrogenation kinetics in $Ti_xNb_{2-x}VCrMnFe$. Kinetic curves at 303 K for (a) $Ti_{1.8}Nb_{0.2}VCrMnFe$ and (b) $Ti_2Nb_0VCrMnFe$ high-entropy alloys. (c) Hydrogenation cyclic test at 303 K for $Ti_2Nb_0VCrMnFe$ for 20 cycles.

Fig. 5 presents the XRD profiles of the HEAs before and after hydrogen absorption for three cycles. Before hydrogenation, all HEAs exhibit the C14 Laves phase, which is well known for its favorable hydrogenation characteristics and kinetics [13,19,25,74]. The corresponding lattice parameters before and after hydrogenation are summarized in Table 2. For the HEAs $Ti_{0.5}Nb_{1.5}VCrMnFe$, TiNbVCrMnFe, and $Ti_{1.5}Nb_{0.5}VCrMnFe$, the storage capacity is comparatively low. Consequently, the XRD peak positions remain largely unchanged after hydrogenation. In contrast, the HEAs $Ti_{1.8}Nb_{0.2}VCrMnFe$ and $Ti_{2}Nb_{0}VCrMnFe$ exhibit the generation of a C14 Laves hydride phase after hydrogenation, resulting in lattice expansion of approximately 10 % [71,75]. Moreover, peak broadening is observed in the XRD pattern for HEAs $Ti_{1.8}Nb_{0.2}VCrMnFe$ and $Ti_{2}Nb_{0}VCrMnFe$, which is attributed to defect generation by lattice expansion and contraction during hydrogen absorption and desorption [10]. However, such lattice defects do not affect the cyclic performance, as confirmed in Fig. 3(f).

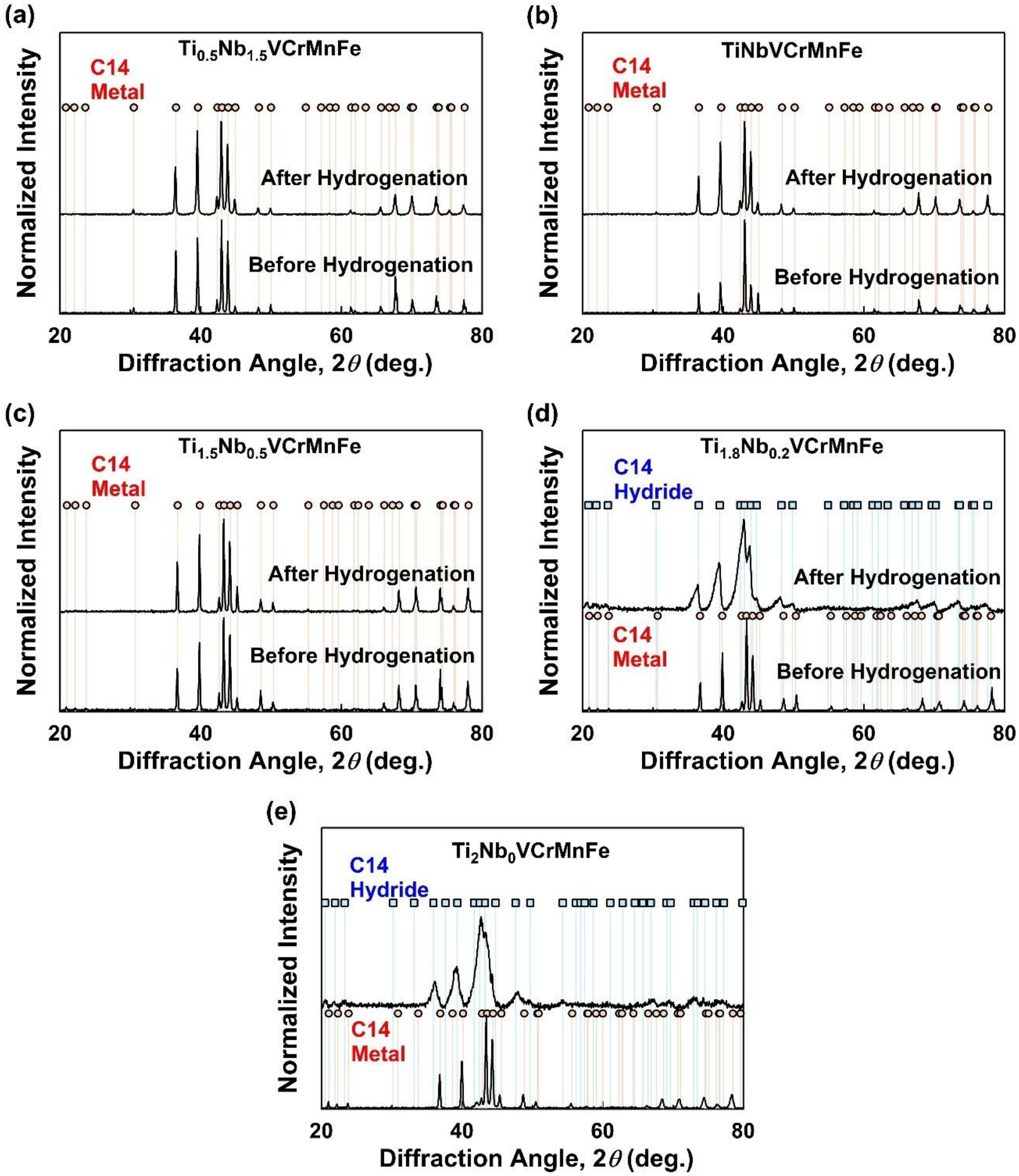

Fig. 5. Transformation from metallic C14 Laves phase to hydride C14 phase in $Ti_xNb_{2-x}VCrMnFe$ with low niobium content. X-ray diffraction (XRD) profile before and after three cycles of hydrogen absorption of (a) $Ti_{0.5}Nb_{1.5}VCrMnFe$, (b) TiNbVCrMnFe, (c) $Ti_{1.5}Nb_{0.5}VCrMnFe$, (d) $Ti_{1.8}Nb_{0.2}VCrMnFe$ and (e) $Ti_2Nb_0VCrMnFe$ high-entropy alloys.

Table 2. Lattice parameters before and after hydrogenation for high-entropy alloy system $Ti_xNb_{2-x}VCrMnFe$ ($x$ = 0.5, 1.0, 1.5, 1.8, 2.0).

| High-entropy alloy | Before hydrogenation | After hydrogenation |
|---|---|---|
| $Ti_{0.5}Nb_{1.5}VCrMnFe$ | $a$ = 4.93 Å, $c$= 8.07 Å | _ |
| TiNbVCrMnFe | $a$ = 4.92 Å, $c$= 8.05 Å | _ |
| $Ti_{1.5}Nb_{0.5}VCrMnFe$ | $a$ = 4.90 Å, $c$= 8.02 Å | _ |
| $Ti_{1.8}Nb_{0.2}VCrMnFe$ | $a$ = 4.89 Å, $c$= 7.99 Å | $a$ = 4.92 Å, $c$= 8.10 Å |
| $Ti_2Nb_0VCrMnFe$ | $a$ = 4.87 Å, $c$= 7.96 Å | $a$ = 5.00 Å, $c$= 8.10 Å |

Fig. 6(a) presents a scanning electron microscopy (SEM) image of $Ti_2Nb_0VCrMnFe$, which exhibits promising hydrogen storage performance at ambient temperature. The associated phase distribution, obtained via electron backscatter diffraction (EBSD), is shown in Fig. 6(b). The EBSD analysis reveals a predominant C14 Laves phase, accompanied by a small amount of BCC phase. The amount of BCC phase calculated through SEM image analysis is 5.2±1.2 vol%. This small amount of the BCC phase was not detected by XRD, possibly because of its small fraction and the superposition of its main peak with the peaks of the C14 phase. While the BCC phase does not have a positive effect on the reversible storage capacity at room temperature due to the high thermal stability of its hydride, its interphase boundaries with the C14 phase are believed to improve the activation and kinetics [76]. SEM image and corresponding EDS elemental maps are displayed in Fig. 6(c), and quantitative EDS analysis is given in Table 3. Proper mixing of all elements in HEA is observed in the EDS mapping. The overall measured composition perfectly fits the nominal composition, as shown in Table 3. EDS analyses confirm that the BCC phase is rich in V. The nanostructures of the $Ti_{1.8}Nb_{0.2}VCrMnFe$ and $Ti_2Nb_0VCrMnFe$ HEAs observed by high-resolution TEM are shown in Fig. 7, confirming the existence of C14 in good consistency with XRD and SEM. Moreover, grain boundaries are also observed, which may contribute to the enhancement of kinetics [8,10,77,78]. Although grain sizes are clearly at the micrometer level in Fig. 7, the quantification of their average sizes was hard due to technical difficulties in polishing them to have a large area for EBSD analysis.

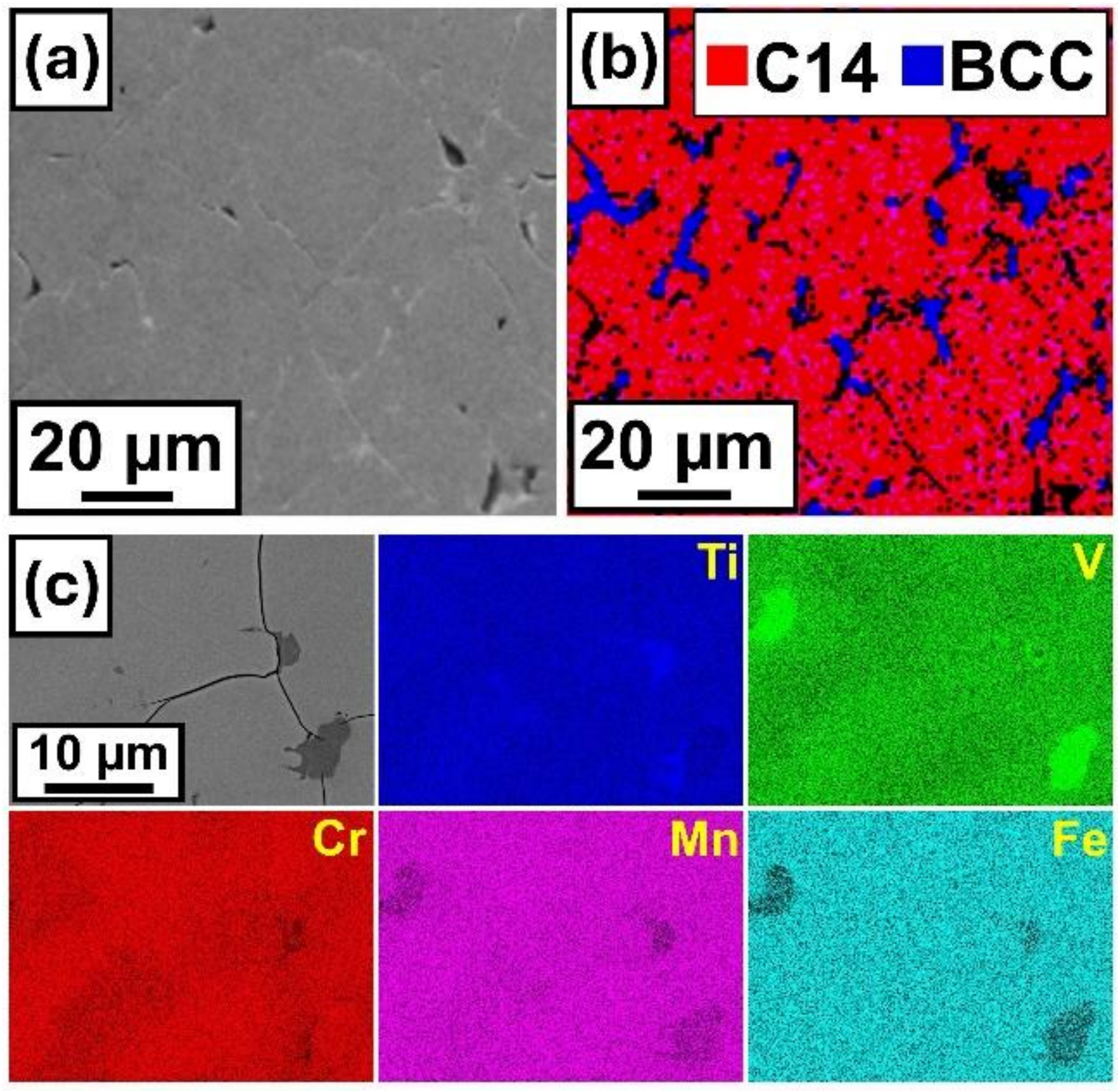

Fig. 6. Existence of C14 phase with a small amount of vanadium-rich BCC phase in niobium-free $Ti_2Nb_0VCrMnFe$ high-entropy alloy. (a) Scanning electron microscopy (SEM) micrograph, (b) corresponding electron backscatter diffraction (EBSD) phase mapping, (c) SEM image and relevant energy dispersive spectroscopy (EDS) elemental mapping for $Ti_2Nb_0VCrMnFe$.

Table 3. Elemental composition of Nb-free $Ti_2Nb_0VCrMnFe$ high-entropy alloy obtained through energy dispersive spectroscopy.

| **$Ti_2Nb_0VCrMnFe$** | | | | |
|---|---|---|---|---|
| **Elements** | **Nominal** | **Measured** | **C14 Laves phase** | **BCC phase** |
| **Ti (at%)** | 33.3 | 33.8 | 32.7 | 33.6 |
| **V (at%)** | 16.6 | 16.4 | 15.3 | 30.9 |
| **Nb (at%)** | 0.0 | 0.0 | 0.0 | 0.0 |
| **Cr (at%)** | 16.6 | 17.5 | 18.6 | 14.5 |
| **Mn (at%)** | 16.6 | 15.3 | 16.1 | 11.1 |
| **Fe (at%)** | 16.6 | 17.0 | 17.3 | 9.9 |

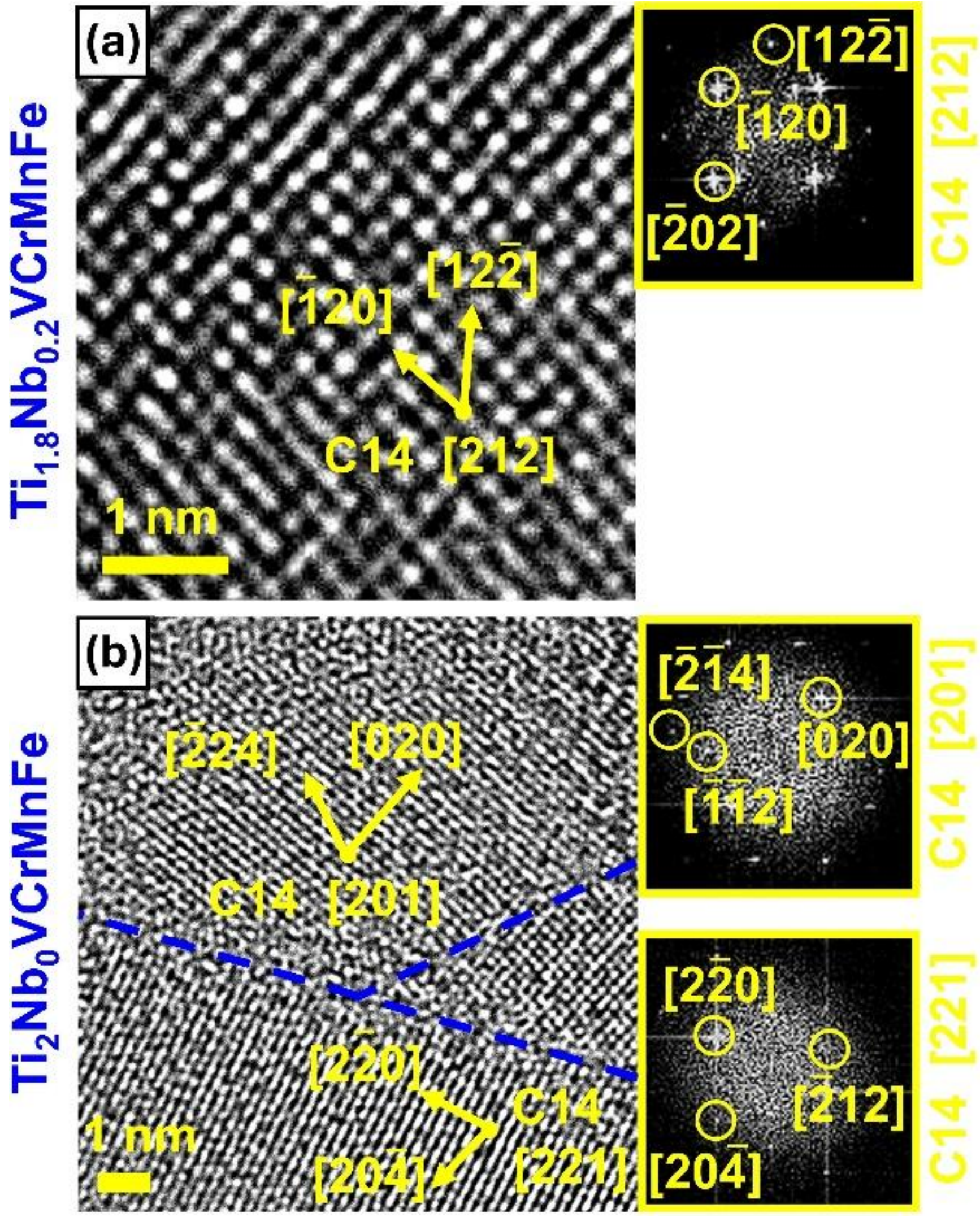

Fig. 7. C14 phase with large fraction of grain boundaries observed in $Ti_xNb_{2-x}$VCrMnFe with low niobium content. Images taken by high-resolution transmission electron microscopy (TEM) and relevant fast Fourier transform (FFT) for (a) $Ti_{1.8}Nb_{0.2}$VCrMnFe and (b) $Ti_2Nb_0$VCrMnFe high-entropy alloys.

**3.2 Machine Learning**

An ML model using ANN was trained using three approaches: first, without removing outliers, second, by removing outliers statistically, and third, by eliminating outliers through a hybrid strategy, as explained in the method section. Fig. 8(a, b) shows the cross-plot of all data without eliminating any outliers through (a) *K*-fold cross-validation and (b) resubstitution validation; Fig. 8(c, d) are cross-plots after statistical outlier removal for (c) *K*-fold cross-validation and (d) resubstitution validation. For the statistical-outlier-removal method, the coefficients of determination are $R^2$ = 0.73 and 0.86 for *K*-fold cross-validation and resubstitution validation, respectively, which are better than those achieved without removing outliers (0.70 for *K*-fold cross-validation and 0.82 for resubstitution validation). The statistical-outlier-removal method shows RMSEs of 5.3 kJ/mol and 3.7 kJ/mol for *K*-fold cross-validation and resubstitution validation, respectively. An improvement in RMSE is also observed after eliminating the outliers, and this validates the reliability of the outlier-detection method employed. A comparison of RMSE using the statistical and hybrid outlier-removal methods indicates slightly better performance for the hybrid method, while the resubstitution validation method always shows better results. The formation enthalpies predicted for the HEA system $Ti_xNb_{2-x}VCrMnFe$ ($x$ = 0, 0.5, 1.0, 1.5, 1.8, 2.0) are shown in Table 4. Statistical outlier removal predicted formation enthalpy of −27.5±1.7 kJ/mol for $Ti_{1.8}Nb_{0.2}VCrMnFe$ and −29.1±1.7 kJ/mol for $Ti_2Nb_0VCrMnFe$. Formation enthalpy predicted by hybrid outlier removal is −24.6±1.4 kJ/mol for $Ti_{1.8}Nb_{0.2}VCrMnFe$ and −26.3±1.0 kJ/mol for $Ti_2Nb_0VCrMnFe$. Although the formation enthalpies predicted by both methods are near the experimental values (−24.9 kJ/mol for $Ti_{1.8}Nb_{0.2}VCrMnFe$ and −25.1 kJ/mol for $Ti_2Nb_0VCrMnFe$), the prediction obtained from the hybrid outlier method is closer to the experimental values. Model performance was also evaluated after excluding data points where a change in entropy was assumed to be −110 J/mol/K. As shown in Fig. S3, the enthalpy prediction does not change for this model, but the $R^2$ value slightly improves to 0.75 and 0.88 after statistical outlier removal for $K$-fold cross-validation and resubstitution validation, respectively. The improvement in the $R^2$ value should be due to some deviations in real entropy compared to the assumption of −110 J/mol/K.

Table 4. Formation enthalpy of high-entropy alloy system $Ti_xNb_{2-x}VCrMnFe$ ($x$ = 0, 0.5, 1.0, 1.5, 1.8, 2.0) predicted through machine learning by artificial neural networks (ANN) using resubstitution validation method after removal of outliers by statistical and hybrid methods in comparison with predictions by Gaussian process regression (GPR) and resubstitution validation methods after removal of outliers using statistical method.

| **High-entropy alloy** | **Formation enthalpy (kJ/mol)** | | |
|---|---|---|---|
| | **ANN (statistical outlier removal)** | **ANN (hybrid outlier removal)** | **GPR** |
| $Ti_0Nb_2VCrMnFe$ | −15.8±4.3 | −13.1±3.2 | −25.1±9.1 |
| $Ti_{0.5}Nb_{1.5}VCrMnFe$ | −19.0±3.2 | −15.5±2.6 | −24.5±7.7 |
| TiNbVCrMnFe | −22.3±1.9 | −18.7±2.5 | −23.6±6.7 |
| $Ti_{1.5}Nb_{0.5}VCrMnFe$ | −25.2±1.7 | −22.2±2.1 | −23.9±5.8 |
| $Ti_{1.8}Nb_{0.2}VCrMnFe$ | −27.5±1.7 | −24.6±1.4 | −25.5±5.5 |
| $Ti_2Nb_0VCrMnFe$ | −29.1±1.7 | −26.3±1.0 | −26.8±5.5 |

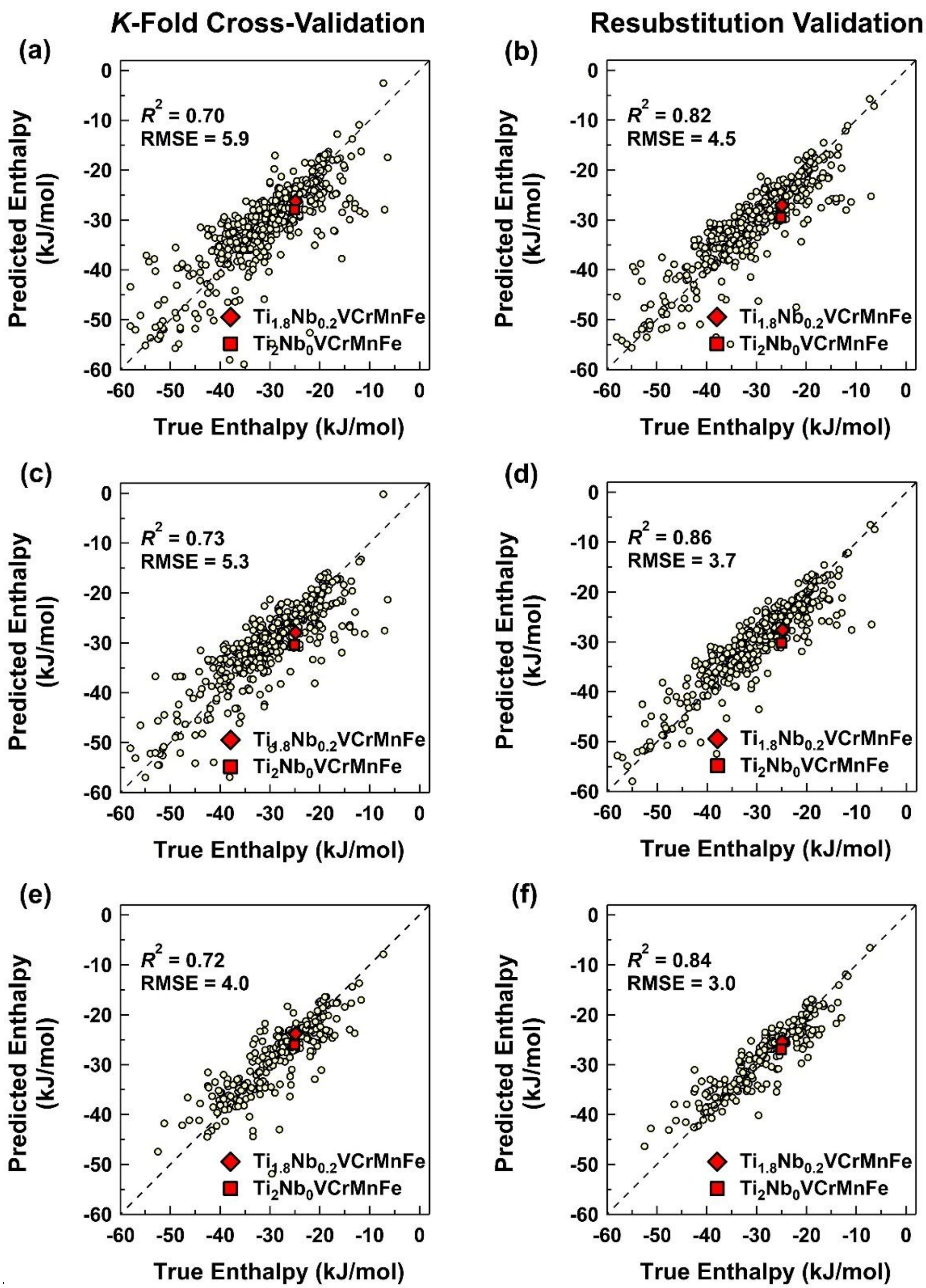

Fig. 8. Removal of outliers through statistical and hybrid methods improved coefficient of determination ($R^2$) and root mean squared error (RMSE). Comparison between predicted enthalpy and true enthalpy for data in Table S1. Machine learning plots of predicted enthalpy data versus true data achieved by artificial neural networks (ANN) using (a, c, e) *K*-fold cross-validation, (b, d, f) resubstitution validation (a, b) without outlier removal, (c, d) after statistical outlier removal and (e, f) after hybrid outlier removal. Data points for $Ti_{1.8}Nb_{0.2}VCrMnFe$ and $Ti_2Nb_0VCrMnFe$ high-entropy alloys were included for comparison.

The formation enthalpy for the HEA system $Ti_xNb_{2-x}VCrMnFe$ ($x$ = 0, 0.5, 1.0, 1.5, 1.8, 2.0) predicted through ANN and GPR is given in Table 4. The ANN results show that increasing Ti content makes the formation enthalpy more negative, which is also observed in experiments as well as in a study on the Ti-Nb alloys [70]. However, GPR predicts an insignificant effect of Ti content on the enthalpy. As will be discussed later, the low reliability of GPR for the HEA system $Ti_xNb_{2-x}VCrMnFe$ system can be attributed to the reversion-to-the-mean feature in GPR when extrapolating the data, whereas ANN does not exhibit this feature [40,79]. It should be noted that ANN predictions are not completely extrapolation as the HEA $Ti_xNb_{2-x}VCrMnFe$ lies well within the training domain of the dataset, as shown by principal component analysis (PCA) and *t*-distributed stochastic neighbor embedding (*t*-SNE) in Fig. S4. However, reversion to the mean in GPR prediction is observed with the presence of the test data away from training data points in Fig. S4(a). Moreover, learning curve analysis, shown in Fig. S5, indicates that the gap between training and cross-validation errors becomes smaller and smaller as it reaches 100% of the data points. Combined with $L_2$ regularization and 5-fold cross-validation, the learning curve confirms that the current dataset of 511 points is sufficient to train the ANN model without significant overfitting.

Partial dependence of enthalpy from different elements is displayed in Fig. 9(a-f). For Ti and V, the formation enthalpy becomes more negative as their content increases, which is characteristic of A-type elements [17]. Whereas, for elements such as Cr, Mn and Fe, the formation enthalpy increases with increasing their content, which is expected as these elements are B-type [17]. Although Nb is an A-type element, its addition makes the enthalpy less negative due to the reduction of the Ti content as a strong A-type element. Sensitivity analysis for the elements Ti, V, Nb, Cr, Mn and Fe is illustrated in Fig. 9(g). In sensitivity analysis, the change in enthalpy is regarded as the impact score when the formation enthalpy of an element is removed from the composition. Therefore, a higher impact score for a corresponding element indicates a greater change in the formation enthalpy. The highest impact score belongs to Fe with −9.6 kJ/mol on the negative side and to Ti with +8.2 kJ/mol on the positive side.

Formation enthalpy for HEA $Ti_xNb_{2-x}VCrMnFe$ ($x$ = 0, 0.5, 1.0, 1.5, 1.8, 2.0) predicted by ANN and GPR using the composition-based models is compared with descriptor-based models, as shown in Fig. 10. ANN predictions trained on chemical compositions and ANN predictions trained on descriptors indicate that increasing Ti content leads to a more negative formation enthalpy. While the GPR model provides good predictions for some compositions, the predicted trends in the whole Ti range diverge from the ANN results. A comparison of *RMSE* without and with using descriptors suggests that it changes from 6.2 kJ/mol to 4.8 kJ/mol for GPR and from 5.3 kJ/mol to 6.8 kJ/mol for ANN. It is observed that the descriptor-based model has a higher accuracy for GPR; however, the ANN model demonstrates a good predictive fidelity using both composition and descriptors.

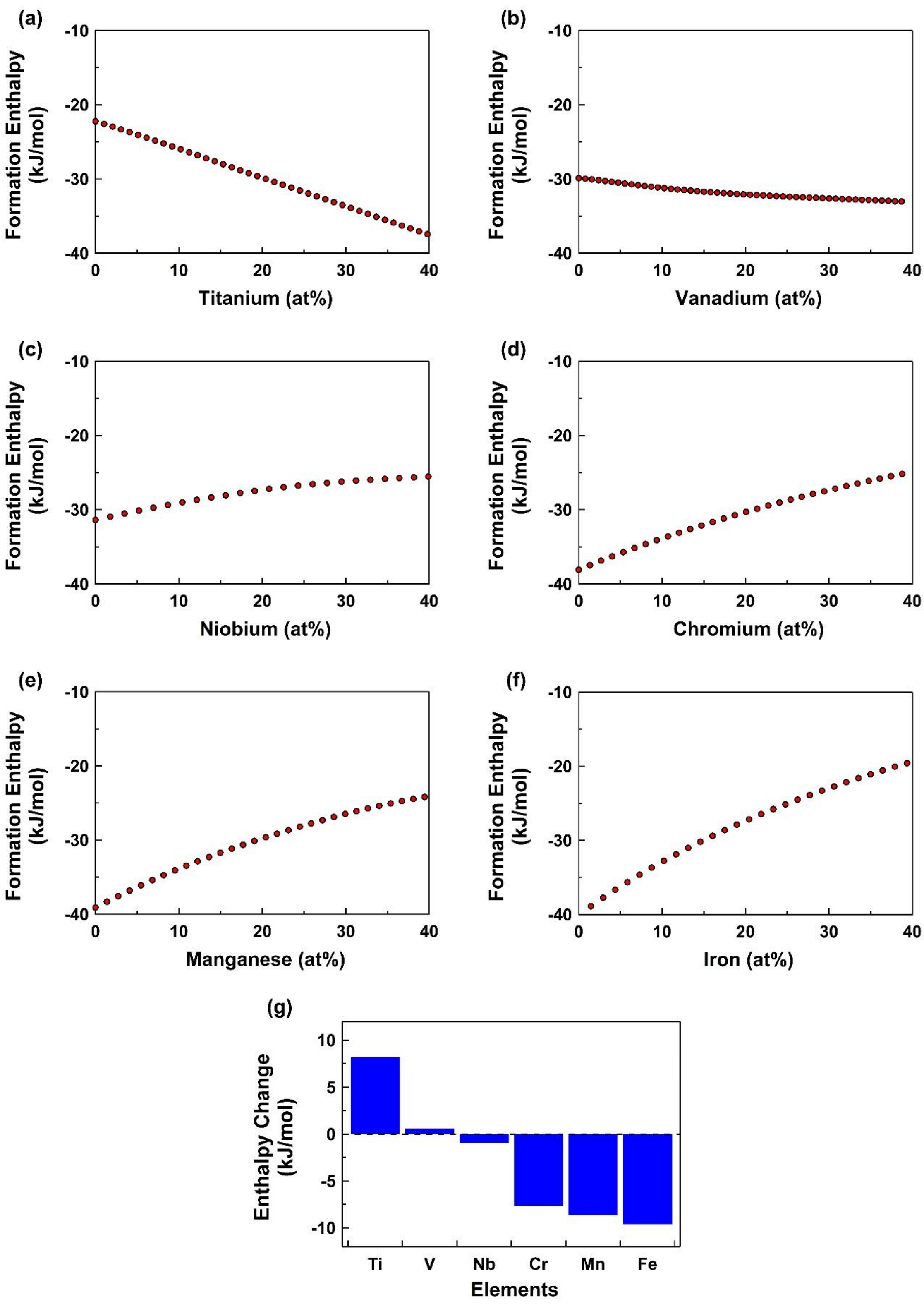

Fig. 9. (a–f) Partial dependence plots (PDPs) of the enthalpies of hydride formation to atomic fractions of titanium, vanadium, niobium, chromium, manganese, and iron. (g) Effect of different elements on enthalpy changes in $Ti_xNb_{2-x}VCrMnFe$ high-entropy alloys.

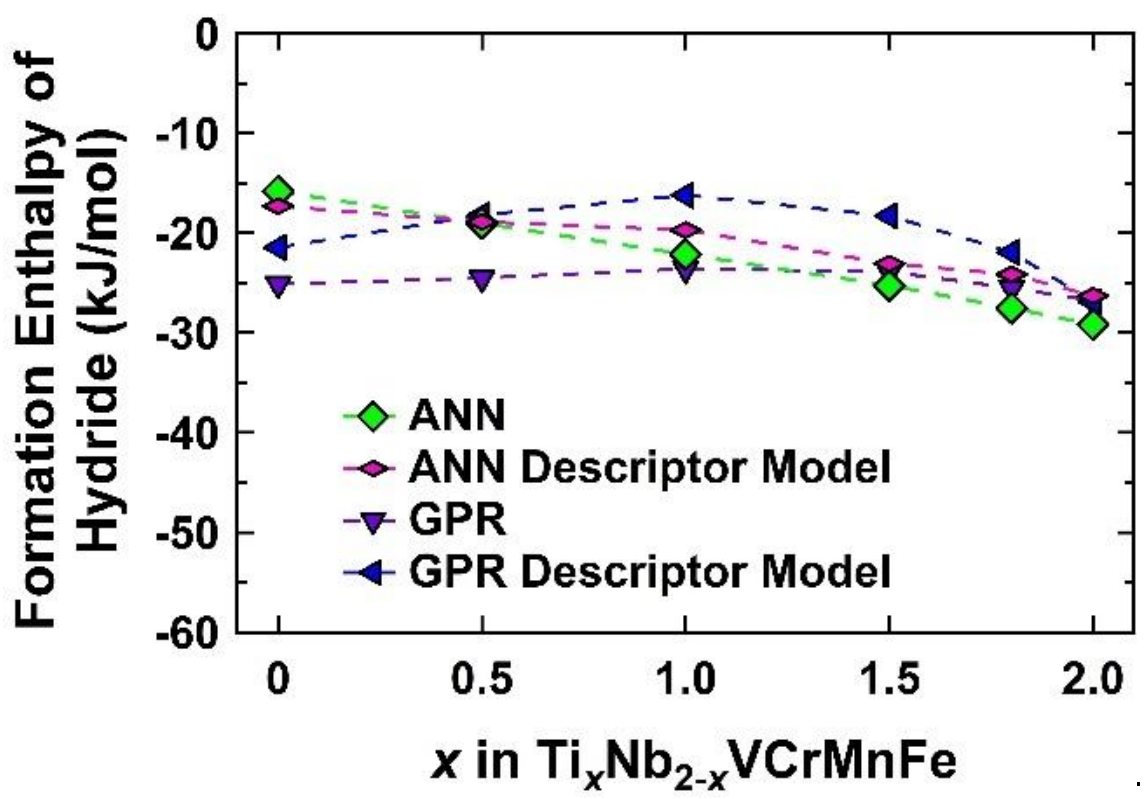

Fig. 10. Comparison of formation enthalpy predicted for high-entropy alloy system $Ti_xNb_{2-x}VCrMnFe$ ($x$ = 0, 0.5, 1.0, 1.5, 1.8, 2.0) by artificial neural network (ANN), Gaussian process regression (GPR) using compositions and descriptor models.

### 3.3 *Ab initio* DFT Calculations

The unit cell volume of the HEAs calculated through *ab initio* DFT calculations is shown in Fig. 11(a) and Table 5, together with experimental data obtained from XRD for comparison. The unit cell volume contracts with rising Ti content both in DFT and experiments, and the difference between experimental data and DFT calculations is less than 4 % (DFT slightly underestimates the volume). This underestimation of volume for DFT can be explained by the approximation of the exchange-correlation functional [68] and is often found in other HEAs [25,80]. Such a small difference confirms the reliability of the modeled superlattices. Lattice contraction by increasing the titanium content empirically suggests that the formation enthalpy of hydride should become more negative. However, metal-hydrogen binding energy has a more significant effect on this thermodynamic parameter (i.e. Ti with an electronegativity of 1.54 makes a stronger binding than Nb with an electronegativity of 1.60 to hydrogen atoms). The formation enthalpy calculated through DFT for the HEA system $Ti_xNb_{2-x}VCrMnFe$ is shown in Fig. 11(b) and Table 5. The formation enthalpy becomes more negative with increasing Ti content, which agrees with the detected reduction in the hydrogenation pressure as Ti concentration increases. DFT results also suggest that the formation enthalpy for the HEAs with $x$ = 0, 0.5, 1.0, 1.5 is +0.2, -8.8, -15.5, -20.3, -32.4 kJ/mol, respectively (Table 5). The enthalpy values for Ti-rich alloys lie within the range for room-temperature hydrogen storage, in good consistency with the PCT isotherms (Fig. 3). These results confirm the successful application of DFT calculations in real-world hydrogen storage systems, as also shown in a few earlier publications [15,25,37,40]. It should be noted that DFT calculations were conducted on the C14 Laves phase, whereas the alloys showed a small amount of BCC phase in the EBSD mapping. This simplification is reasonable because EDS results reveal that since the fraction of the BCC phase is minor, the composition of the C14 phase is similar to the nominal composition of the alloys (Table 3).

This trend highlights the critical role of the Ti/Nb ratio in HEAs for hydrogen storage. With increasing the fraction of Nb, having a larger atomic size, the lattice is expanded, as shown in Fig. 11(a). Lattice expansion is empirically considered effective to destabilize hydrides by providing a larger lattice space for releasing hydrogen. The destabilization of hydride by decreasing the Ti/Nb fraction is evident from Fig. 11(b), where the hydride formation enthalpy becomes less negative with increasing the Nb content. These results confirm that changing the Ti/Nb ratio in HEAs can be used as a strategy to tune the plateau pressure for solid-state hydrogen storage.

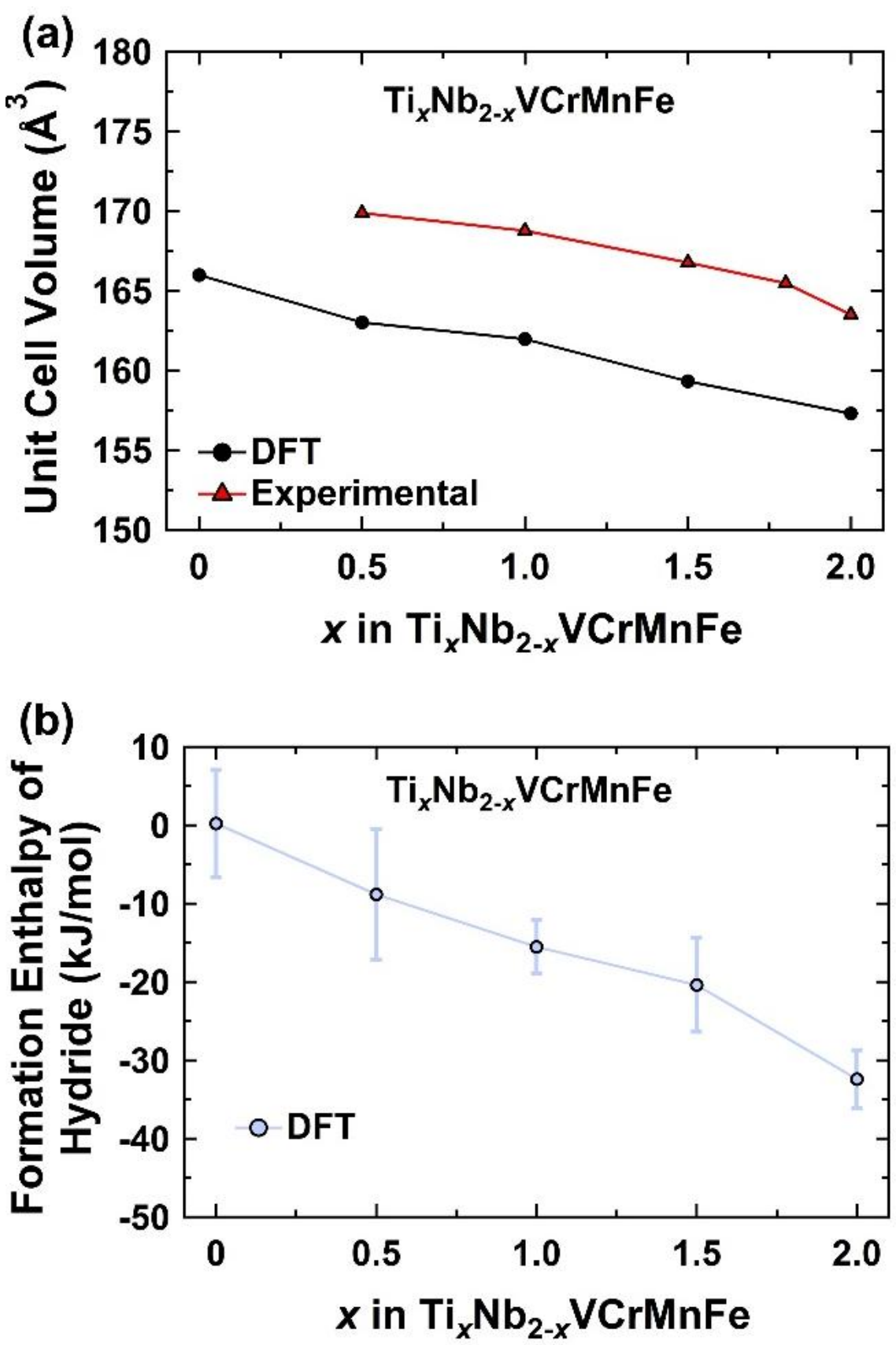

Fig. 11. Contraction of volume and reduction of hydride formation enthalpy for $Ti_xNb_{2-x}$VCrMnFe with increasing titanium content. (a) Unit cell volume of non-hydrogenated $Ti_xNb_{2-x}$VCrMnFe calculated in DFT as a function of titanium content, together with experimental data. (b) Hydride formation enthalpy versus titanium content in $Ti_xNb_{2-x}$VCrMnFe high-entropy alloys. Error bars in (b) indicate range of maximum to minimum enthalpy.

Table 5. Formation enthalpy of high-entropy alloy system $Ti_xNb_{2-x}$VCrMnFe ($x$ = 0, 0.5, 1.0, 1.5, 2.0) calculated through DFT calculations.

| **High-entropy alloy** | **Unit cell volume of HEA (Å$^3$)** | **Formation enthalpy calculated through DFT (kJ/mol)** |
|---|---|---|
| $Ti_0Nb_2$VCrMnFe | 166.0 | 0.2 |
| $Ti_{0.5}Nb_{1.5}$VCrMnFe | 163.0 | −8.8 |
| TiNbVCrMnFe | 162.0 | −15.5 |
| $Ti_{1.5}Nb_{0.5}$VCrMnFe | 159.3 | −20.3 |
| $Ti_2Nb_0$VCrMnFe | 157.3 | −32.4 |

## 4. Discussion

Design of HEAs with favorable thermodynamics and kinetics remains a challenge. Common design methods include experiments, *ab initio* calculations and machine learning. All these methods have their own advantages and disadvantages. For example, experimental results provide valuable insights in real conditions; however, they require both financial and time resources. *Ab initio* calculations using DFT provide insight into fundamental electronic structure and thermodynamic properties, but they demand a lot of computational resources and the results rely on approximations for the exchange and correlation energy. ML is useful in predicting properties that are difficult for the human mind to comprehend, but the selection of appropriate data can be challenging. However, these techniques provide reliable insights and can be used together to complement each other and thereby provide a deeper understanding of materials science. This study introduces novel HEAs for solid-state hydrogen storage with a general composition $Ti_xNb_{2-x}VCrMnFe$ ($x$ = 0.5, 1.0, 1.5, 1.8, 2.0). The formation enthalpy of HEAs, calculated using experiments, DFT and ML through ANN, shows good agreement, as illustrated in Fig. 12. A trend is observed, using DFT and ML methods, that with increasing Ti content or reducing Nb content, the formation enthalpy becomes more negative. Although it is technically hard to quantitatively show the experimental trends within all this compositional range (due to pressure limits in the authors' gas absorption facility), the PCT data in Fig. 3(f) agree with the trends in Fig. 12. Moreover, the absolute enthalpy values estimated using DFT and ML are reasonably close to each other and to experimental values for Ti-rich compounds. The following questions in this study need to be further discussed. (i) Why is there a slight discrepancy between ML results and DFT results? (ii) Why is the ML prediction through ANN better than GPR for the HEA system $Ti_xNb_{2-x}VCrMnFe$ ($x$ = 0.5, 1.0, 1.5, 1.8, 2.0)? (iii) Nb is considered an A-type element, but why does it increase the enthalpy like a B-type element? (iv) How do current HEAs compare with existing HEAs?

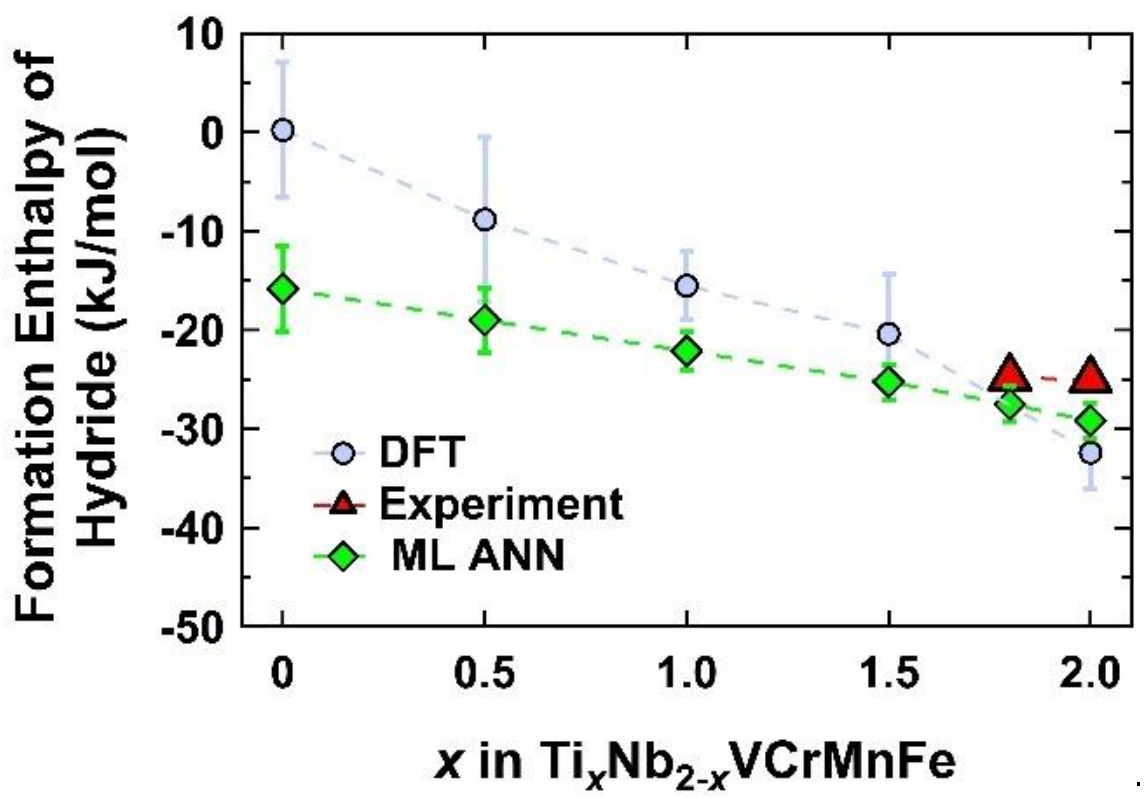

Fig. 12. Comparing hydride formation enthalpy for high-entropy alloy system $Ti_xNb_{2-x}VCrMnFe$ ($x$ = 0, 0.5, 1.0, 1.5, 1.8, 2.0) obtained by experiments, *ab initio* calculations via density functional theory (DFT), and machine learning (ML) through artificial neural networks (ANN).

Regarding the first question, the formation enthalpy through ML and DFT shows a larger discrepancy in the Nb-rich region, whereas this becomes lower in the Nb-poor region. The data available in the literature based on Nb-type alloys with C14 Laves phase are very limited, as these alloys mostly form the BCC phase [28-31,70]. So, ML predictions in Nb-rich regions may not be as accurate as in Nb-poor regions. Moreover, DFT calculation is conducted for a single C14 Laves phase, whereas in experiments, there are small amounts of other phases, which may influence the formation enthalpy. Additionally, DFT calculation assumes a fixed H/M ratio with full hydrogen loading, while ML is trained on experimental data, which may slightly deviate from ideal hydrogen loading. Therefore, the combined effect of limited training data in the Nb-rich area and the presence of dual phases in the experiment, in contrast to the single-phase calculations in DFT, can lead to the observed discrepancy between ML predictions and DFT calculations.

Regarding the second question, the formation enthalpy predicted by ANN shows the same trend as experimental and DFT results, whereas for GPR predictions, the formation enthalpy first slightly increases and then slightly decreases, as shown in Table 4. The prediction through GPR for Nb-poor HEAs is consistent with DFT and experiments, while the error appears with increasing the Nb content. Since Nb-type alloys with C14 Laves phase are limited [28-31,70], the Nb-rich alloys should be evaluated by extrapolation. GPR is a kernel-based approach and performs very well for interpolation, as shown earlier [15,40,42,47]. However, GPR is based on probabilistic models that revert to prior mean in outside domain regions, thereby leading to poor extrapolation [40,79]. In contrast, ANN is based on deterministic function approximators and does not exhibit reversion to the mean [40,79]. Therefore, the trend of decreasing formation enthalpy with decreasing the Nb content can be correctly predicted by ANN.

Regarding the third question, Nb is considered an A-type element for its ability to form hydrides [76]. Being an A-type element, increasing the Nb content should make the hydride more stable, i.e. the formation enthalpy should become more negative. However, the predictions through ML and DFT show a reverse trend, and experimental PCT isotherms in Fig. 3 also indicate that the plateau pressure rises with rising the Nb content. PDP of Nb, as presented in Fig. 9(c), shows an unusual positive slope, confirming the behavior of Nb like a B-type element. Additional sensitivity analysis for Nb, as shown in Fig. 9(g), also confirms that the enthalpy changes to positive values by adding Nb. This behavior of Nb, which is consistently observed using experiments, DFT and ML, should not be considered individually, as its addition is accompanied by the reduction of Ti as a strong A-type element (the formation enthalpy of titanium hydride is −67 kJ/mol, while it is only −20 kJ/mol for niobium hydride [70]). Therefore, the enhancement of the Nb/Ti ratio should increase the enthalpy, a fact that was correctly predicted by ANN despite limited data about Nb-containing C14 materials.

Regarding the fourth point of discussion, Table 6 shows the formation enthalpy of some HEAs along with their reversible storage capacity at ambient temperature [13,15,19,73,81]. Two HEAs, $Ti_2Nb_0VCrMnFe$ and $Ti_{1.8}Nb_{0.2}VCrMnFe$, with a reversible storage capacity of 1.5 wt%, exhibit reasonable capacity compared to the reported HEAs and stand among HEAs that can absorb/desorb hydrogen at room temperature. This capacity is comparable to commercial $LaNi_5$, but the current alloys are cheaper than $LaNi_5$. Many other HEAs still require high temperatures for the desorption of hydrogen. For example, although some HEAs, such as $V_{35}Ti_{35}Cr_{10}Fe_6Mn_{14}$, with the BCC structure, absorb up to 3.79 wt%, their reversibility at room temperature is very low [28].

In summary, this study confirms the high potential of ANN in predicting the enthalpy of formation of a specific system of high-entropy hydrides. While this study concentrated on a Ti-V-Nb-Cr-Mn-Fe system, in which Nb with a large atomic diameter was added to the A site, future studies can consider adding an element with a large atomic size, like Mo, in the B site. Moreover, expanding the composition on both A and B sites can further clarify the applicability of ANN for the discovery of HEAs for hydrogen storage at room temperature. Finally, the authors believe that ANN has high potential for some other hydrogen-related applications, such as hydrogen production, as high-entropy materials have received attention in such applications in recent years, particularly by the group of current authors [82,83].

Table 6. Formation enthalpy and hydrogen storage capacity of some high-entropy alloys with capability for reversible hydrogen storage at room temperature.

| High-entropy alloy | Hydrogen storage capacity (wt%) | Formation enthalpy (kJ/mol) | Reference |
|---|---|---|---|
| $Ti_2Nb_0VCrMnFe$ | 1.5 | −25.1 | This work |
| $Ti_{1.8}Nb_{0.2}VCrMnFe$ | 1.5 | −24.9 | This work |
| $TiV_2ZrCrMnFeNi$ | 1.6 | - | [19] |
| $TiV_{1.5}ZrCr_{0.5}MnFeNi$ | 1.2 | - | [13] |
| $TiV_{1.5}Zr_{1.5}CrMnFeNi$ | 1.8 | - | [13] |
| TiZrCrMnFeNi | 1.6 | −28.8 | [15] |
| $Ti_{0.5}Zr_{1.5}CrMnFeNi$ | 1.6 | −36.5 | [15] |
| $Ti_{20}Zr_{20}Mn_{20}Fe_{20}Co_{20}$ | 1.52 | −32.4 | [73] |
| $Ti_{15}Zr_{22}Mn_{27}Fe_{26}Co_{10}$ | 1.5 | −32.5 | [73] |
| $Ti_{1.1}Mn_1Cr_1$ | 1.61 | −22.9 | [81] |
| $Ti_{0.99}Zr_{0.11}Mn_1Cr_1$ | 1.68 | −25.5 | [81] |
| $Ti_{0.935}Zr_{0.165}Mn_1Cr_1$ | 1.81 | −26.6 | [81] |
| $Ti_{0.88}Zr_{0.22}Mn_1Cr_1$ | 1.90 | −28.3 | [81] |
| $Ti_{0.935}Zr_{0.165}Mn_1Cr_{0.95}Mo_{0.05}$ | 1.88 | −26.2 | [81] |
| $Ti_{0.935}Zr_{0.165}Mn_1Cr_{0.9}Mo_{0.1}$ | 1.78 | −23.7 | [81] |
| $Ti_{0.935}Zr_{0.165}Mn_1Cr_{0.85}Mo_{0.15}$ | 1.67 | −21.7 | [81] |
| $Ti_{0.935}Zr_{0.165}Mn_1Cr_{0.98}W_{0.02}$ | 1.52 | −26.3 | [81] |
| $Ti_{0.935}Zr_{0.165}Mn_1Cr_{0.95}W_{0.05}$ | 1.43 | −24.3 | [81] |
| $Ti_{0.935}Zr_{0.165}Mn_1Cr_{0.9}W_{0.1}$ | 1.36 | −22.6 | [81] |

## 5. Conclusions

This study employs machine learning via artificial neural networks (ANN) coupled with density functional theory (DFT) and experiments for thermodynamic optimization of high-entropy alloys for hydrogen storage at ambient temperature. A new high-entropy system with composition $Ti_xNb_{2-x}VCrMnFe$ ($x$ = 0.5, 1.0, 1.5, 1.8, 2.0) for storing hydrogen at ambient temperature is introduced. The hydride formation enthalpy was estimated by training ANN on available literature data, and the results were compared with those from DFT and experimental measurements. A consistent trend in formation enthalpy is observed with the Ti/Nb content using three different methods: the hydride formation enthalpy became less negative by increasing the Nb content. The predicted values are also quantitatively close using the three methods, confirming a good precision of ANN in designing hydrogen storage materials.

**CRediT Authorship Contribution Statement**

All authors: Conceptualization, Methodology, Investigation, Validation, Writing – review & editing.

**Declaration of Competing Interest**

All authors declare no competing financial interests or personal relationships that can affect the research presented in this manuscript.

**Data availability**

All the data used in the current manuscript are made available through the request to the corresponding author.

**Acknowledgment**

The author Shivam Dangwal acknowledges the MEXT, Japan, for a scholarship. This study is partly supported by the ASPIRE project of the Japan Science and Technology Agency (JST) (JPMJAP2332). Pranav Kumar and Yuji Ikeda acknowledge the Deutsche Forschungsgemeinschaft (DFG, German Research Foundation), project number 519607530. Blazej Grabowski acknowledges funding from the European Research Council (ERC) under the European Union's Horizon 2020 research and innovation programme (grant agreement No. 865855). Blazej Grabowski also acknowledges support by the State of Baden–Württemberg through bwHPC and the DFG through grant no INST 40/575-1 FUGG (JUSTUS 2 cluster) and the SFB1333 (project ID 358283783-CRC 1333/2 2022).

**Appendix**

Artificial neural network model specifications, dimensionality reduction analysis and learning curve analysis are given in the Supplementary Material. The dataset for hydrogen-storing alloys is also given in Supplementary Material Table S1. These data were taken from [70-72,81,84-189]. Formation enthalpy was not given explicitly in reference [106,112-113,115-117,120-121, 135,149-151,158], so the van 't Hoff method was utilized for estimating formation enthalpy using PCT isotherms, assuming the entropy changes are −110 J/mol/K [31-33]. This assumption for the entropy is reasonable, as a similar value has been reported in several publications [81,85,86,89,90,92,94,97,100,108].